\documentclass[traditabstract]{aa} 

\usepackage[fleqn]{amsmath}
\usepackage{amssymb}
\usepackage{textcomp}
\usepackage{graphicx}
\usepackage{txfonts}
\usepackage{natbib}
\usepackage{url}
\usepackage{multirow}
\usepackage{subfigure}
\usepackage[usenames,dvipsnames]{xcolor}
\usepackage{bm}
\usepackage[version=4]{mhchem}

\newcommand{\nzdp}{{\rm N^{15}ND^{+}}}
\newcommand{\zndp}{{\rm ^{15}NND^{+}}}

\begin{document}

\title{Depletion of $^{15}$N in the center of L1544: Early transition from atomic to molecular nitrogen?}
\titlerunning{Depletion of $^{15}$N in L1544}
\authorrunning{Furuya et al.}

\author{K. Furuya\inst{\ref{inst1}} 
\and Y. Watanabe\inst{\ref{inst2},\ref{inst3}}
\and T. Sakai\inst{\ref{inst4}}
\and Y. Aikawa\inst{\ref{inst5}}
\and S. Yamamoto\inst{\ref{inst6}}
}

\institute{Center for Computational Sciences, University of Tsukuba, 1-1-1 Tennoudai, 305-8577 Tsukuba, Japan\\
\email{furuya@ccs.tsukuba.ac.jp}\label{inst1} 
\and  Division of Physics, Faculty of Pure and Applied Sciences, University of Tsukuba, Tsukuba, Ibaraki 305-8571, Japan  \label{inst2}
\and Tomonaga Center for the History of the Universe, Faculty of Pure and Applied Sciences,
University of Tsukuba, Tsukuba, Ibaraki 305-8571, Japan \label{inst3}
\and Graduate School of Informatics and Engineering, The University of Electro-
Communications, Chofu, Tokyo 182-8585, Japan \label{inst4}
\and Department of Astronomy, The University of Tokyo, Bunkyo-ku, Tokyo 113-0033, Japan \label{inst5}
\and Department of Physics, The University of Tokyo, Bunkyo-ku, Tokyo 113-0033, Japan \label{inst6}
}


 
\abstract
{
We performed sensitive observations of the $\nzdp$(1--0) and $\zndp$(1--0) lines 
toward the prestellar core L1544 using the IRAM 30m telescope.
The lines are not detected down to 3$\sigma$ levels in 0.2 km s$^{-1}$ channels of $\sim$6 mK.
The non-detection provides the lower limit of the $^{14}$N/$^{15}$N ratio for \ce{N2D+} of $\sim$700-800, 
which is much higher than the elemental abundance ratio in the local ISM of  $\sim$200-300. 
The result indicates that \ce{N2} is depleted in $^{15}$N in the central part of L1544, 
because \ce{N2D+} preferentially traces the cold dense gas, and because it is a daughter molecule of \ce{N2}.
In-situ chemistry is unlikely responsible for the $^{15}$N depletion in \ce{N2};
neither low-temperature gas phase chemistry nor isotope selective photodissociation of \ce{N2} explains the $^{15}$N depletion; 
the former prefers transferring $^{15}$N to \ce{N2}, 
while the latter requires the penetration of interstellar FUV photons into the core center.
The most likely explanation is that $^{15}$N is preferentially partitioned into ices compared to  $^{14}$N 
via the combination of isotope selective photodissociation of \ce{N2} 
and grain surface chemistry in the parent cloud of L1544 or in the outer regions of L1544 which are not fully shielded from the interstellar FUV radiation.
The mechanism is the most efficient at the chemical transition from atomic to molecular nitrogen. 
In other words, our result suggests that the gas in the central part of L1544 already went trough the transition from atomic to molecular nitrogen 
in the earlier evolutionary stage, and that \ce{N2} is currently the primary form of gas-phase nitrogen.

}


\maketitle

\section{Introduction}
\label{sec:intro}
Nitrogen is the fifth most abundant element in the universe.
Our understanding of nitrogen chemistry in star-forming regions is limited compared to other volatile elements, such as carbon and oxygen.
The dominant form of gaseous nitrogen in star-forming regions is unclear \citep[e.g.,][]{bergin07}, 
while it is theoretically expected to be either atomic N or \ce{N2} \citep[e.g.,][]{aikawa05,legal14}.
One of the main limiting factors is that neither atomic N nor \ce{N2} is directly observable in the cold and dense gas of star-forming regions. 
However, the \ce{N2} abundance can be constrained indirectly by observing a proxy molecule \ce{N2H+} \citep[e.g.,][]{maret06}.

The observational and theoretical studies of nitrogen isotope fractionation in star-forming regions can help to constrain nitrogen chemistry.
Nitrogen has two stable isotopes, $^{14}$N and $^{15}$N.
The elemental abundance ratio [$^{14}$N/$^{15}$N]$_{\rm elem}$ in the local interstellar medium (ISM) has been estimated 
to be $\sim$200-300 from the absorption line observations of N-bearing molecules toward diffuse clouds \citep{lucas98,ritchey15}.
L1544 is a prototypical prestellar core located in the Taurus molecular cloud complex.
In L1544, the $^{14}$N/$^{15}$N ratio of several different molecules has been measured:
$^{14}$\ce{N2H+}/N$^{15}$NH$^+$ = $920^{+300}_{-200}$, $^{14}$\ce{N2H+}/$^{15}$NNH$^+$ = $1000^{+260}_{-220}$ \citep{bizzocchi13,redaelli18}, 
\ce{NH2D}/$^{15}$\ce{NH2D} $>$ 700  \citep{gerin09}, 
\ce{CN}/C$^{15}$N = $500 \pm 75$ \citep{hilyblant13b}, and  \ce{HCN}/HC$^{15}$N = 257 \citep{hilyblant13a}.
Among the measurements, the significant depletion of $^{15}$N in \ce{N2H+} is the most challenging for the theory of $^{15}$N fractionation.
In general, molecules formed at low temperatures are enriched in $^{15}$N through gas-phase chemistry 
triggered by isotope exchange reactions \citep[e.g.,][]{terzieva00}.
A $^{15}$N-bearing molecule has a slightly lower zero-point energy compared to the corresponding $^{14}$N isotopologue.
This results in endothermicity for the exchange of $^{15}$N for $^{14}$N, which inhibits this exchange 
at low temperature enabling the concentration of $^{15}$N in molecules.
Indeed, astrochemical models for prestellar cores, which consider a set of nitrogen isotope exchange reactions, 
have predicted that atomic N is depleted in $^{15}$N, while \ce{N2} (and thus \ce{N2H+}) is enriched in $^{15}$N \citep[e.g.,][]{charnley02}.
The model prediction clearly contradicts with the observation of the \ce{N2H+} isotopologues in L1544.
The $^{15}$N depletion in \ce{N2H+} was recently found in other prestellar cores, L183, L429, and L694-2, as well \citep{redaelli18}.
Furthermore, \citet{roueff15} recently pointed out the presence of activation barriers for some key nitrogen isotope exchange reactions, 
based on their quantum chemical calculations.
Then $^{15}$N fractionation triggered by isotope exchange reactions may be much less efficient 
than had been previously thought \citep[][but see also Wirstr\"om \& Charnley 2018]{roueff15}.


Another mechanism that can cause $^{15}$N fractionation is photodissociation of \ce{N2} \citep{heays14}.
\ce{N2} photodissociation is prone to self-shielding.
Because $^{14}$N$^{15}$N is much less abundant than $^{14}$\ce{N2}, $^{14}$N$^{15}$N needs a higher column density of the ISM gas for self-shielding.
This makes \ce{N2} photodissociation an isotope selective process.
As a result, \ce{N2} is depleted in $^{15}$N, which is consistent with the observation of the \ce{N2H+} isotopologues in L1544.
But, isotope selective photodissociation of \ce{N2} is efficient only for the limited regions, 
where the interstellar UV radiation field is not significantly attenuated \citep{heays14,furuya18a}.
The prestellar core L1544 has high density and $A_V$ ($>$10 mag for a mm dust continuum peak).
Detection of carbon chain species, such as \ce{C3H2}, however, may indicate that the interstellar UV radiation penetrates to
moderate depth in L1544 \citep[][]{spezzano16}.
Then it is unclear whether the isotope selective photodissociation of \ce{N2} is at work in L1544, and how it affects the measurement of 
the $^{14}$N/$^{15}$N ratio of \ce{N2H+}.

In order to test the selective photodissociation scenario, we observe $^{15}$N isotopologues of \ce{N2D+} toward prestellar core L1544.
Compared with \ce{N2H+}, \ce{N2D+} selectively traces colder and denser regions (i.e., core center) \citep{caselli02a}, 
where the attenuation of the interstellar UV radiation field is more significant.
If the isotope selective photodissociation of \ce{N2} by the penetrating UV radiation is the cause of the $^{15}$N depletion in \ce{N2H+}, 
the $^{14}$N/$^{15}$N ratio of \ce{N2D+} should be smaller than that of \ce{N2H+} and be close to [$^{14}$N/$^{15}$N]$_{\rm elem}$. 
Moreover, \ce{N2D+} is less optically thick than \ce{N2H+}, 
which allows us to accurately evaluate the column density of the $^{14}$N isotopologue and thus the $^{14}$N/$^{15}$N ratio, 
although more sensitive observations are required for the detection of the $^{15}$N isotopologues of \ce{N2D+} than those of \ce{N2H+}. 

\section{Observations}
\label{sec:obs}
We observed the N$^{15}$ND$^+$(1--0), $^{15}$NND$^+$(1--0), and N$_2$D$^+$(1--0) transitions toward 
the prestellar core L1544 with the IRAM 30m telescope at Pico Veleta on 2017 December 22-24.
We tracked the L1544 continuum dust emission peak at 1.3 mm, where $^{15}$N isotopologues of \ce{N2H+} were previously detected \citep{bizzocchi10,bizzocchi13}.
The observed position is ($\alpha_{\rm J2000}$, $\sigma_{\rm J2000}$) =  ($05^{\rm h}04^{\rm m}17^{\rm s}.21$, $25^{\circ}10'42''.8$) \citep{caselli02a}.
The telescope pointing was checked every two hours by observing the continuum source 0439+360 near the target position and was assured to be better than $\pm 3''$.
The half beam power width was $32''-33''$.

We employed Eight Mixer Receiver (EMIR) E090 with dual polarization mode.  
The system noise temperatures were typically from 70 K to 130 K during the observation run.
The N$^{15}$ND$^+$(1--0) and $^{15}$NND$^+$(1--0) transitions were observed simultaneously (Set 1), 
while the N$_2$D$^+$(1--0) transition was observed with a different frequency setting (Set 2).
The hyperfine components and their relative intensities of the $\nzdp$(1--0) and $\zndp$(1--0) transitions were experimentally studied by \citet{dore09}, 
and they are listed in Table \ref{table:freq} in the Appendix.
A frequency-switching mode was employed with a frequency offset of 7.35 MHz.
We used eight Fourier transform spectrometers (FTS) autocorrelators with the bandwidth of 1820 MHz.
The frequency resolution of 50 kHz corresponds to 0.2 km s$^{-1}$ at 75 GHz.
We integrated the spectrum for a total on-source time of 5.2 hr for Set 1 and 0.4 hr for Set 2. 
Two orthogonal polarizations were simultaneously observed, and are averaged together to produce the final spectrum.
The main beam temperature ($T_{\rm MB}$) is derived by $T_{\rm a}^* F_{\rm eff}/B_{\rm eff}$, 
where $T_{\rm a}^*$ is the antenna temperature, $F_{\rm eff}$ is the forward efficiency (95 \%), 
and $B_{\rm eff}$ is the main beam efficiency (74 \%).
The final rms noise is 2.3 mK in $T_{\rm MB}$ for $\nzdp$(1--0), 2.1 mK for $\zndp$(1--0), and 11 mK for N$_2$D$^+$(1--0).
The N$^{15}$NH$^+$(1--0) and $^{15}$NNH$^+$(1--0) transitions were also observed in Set 1.
Both transitions were detected, and the obtained spectra are similar to those obtained in the framework of ASAI IRAM 30m large program \citep{desimone18,lefloch18}, 
who observed the same object and the same position with the same velocity resolution, but employing Wobbler Switching mode.
We do not discuss the observations of the $^{15}$N isotopologues of \ce{N2H+} in this work, 
because they are studied in detail in previous work \citep{bizzocchi10,bizzocchi13,desimone18}.

\section{Results}
\label{sec:result}
\begin{figure}
\resizebox{\hsize}{!}{\includegraphics{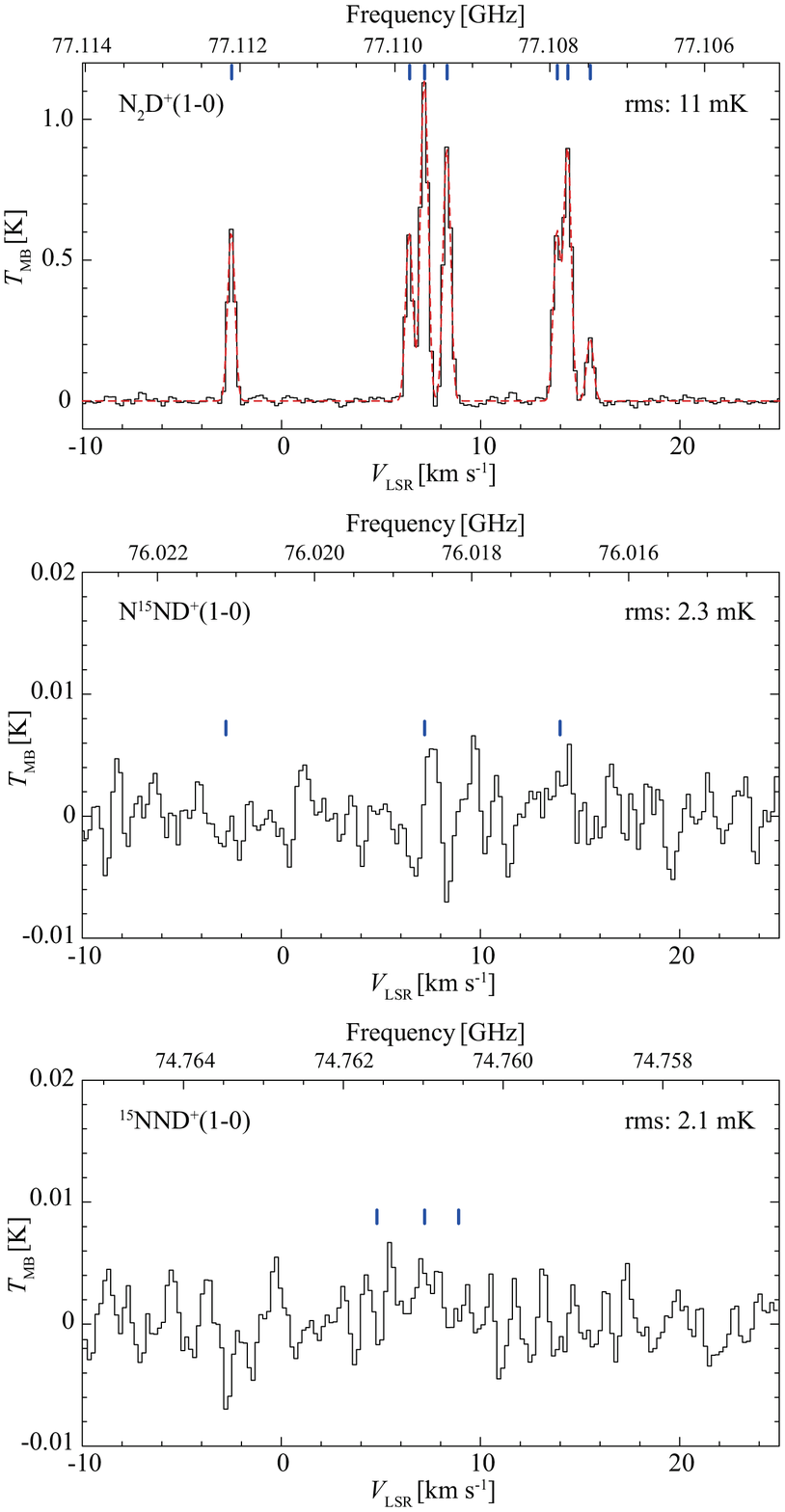}}
\caption{
Spectra of the \ce{N2D+}(1--0) transition (top panel), the $\nzdp$(1--0) transition (middle panel), and the $\zndp$(1--0) transition (bottom panel) 
observed toward L1544.
The intensity scale is the main beam temperature.
In the top panel, the red curve depicts the result of the HFS fit.
The $\nzdp$(1--0) transition and the $\zndp$(1--0) transition were not detected  down to 3$\sigma$ levels in 
0.2 km s$^{-1}$ channels of 6.9 mK and 6.3 mK, respectively.
The vertical blue lines indicate the positions of the expected hyperfine components.
}
\label{fig:spect}
\end{figure}

The data were processed by using the GILDAS software \citep{pety05}.
The \ce{N2D+}(1--0) transition was clearly detected, 
while N$^{15}$ND$^+$(1--0) and $^{15}$NND$^+$(1--0) were not detected, as shown in Figure \ref{fig:spect}.
Line parameters for \ce{N2D+}(1--0) are estimated by using the HFS routine implemented in CLASS.
The derived total optical depth of the lines and the FWHM linewidth are $3.08 \pm 0.19$  and $0.406 \pm 0.003$ km s$^{-1}$, respectively.
The main component of the \ce{N2D+}(1--0) transition (77.1096162 GHz), which has a fraction of 7/27 of the total line strength, is 
marginally optically thick ($\sim$0.8).
We derive the total column density of \ce{N2D+} using Equation (A1) of \citet{caselli02b},
which is valid for optically thick emission.
For the column density calculation, the excitation temperature ($T_{\rm ex}$) is set to be 5 K, which was previously derived from 
\ce{N2H+}(1--0) and \ce{N2D+}(2--1) observations toward the same object and the same position \citep{caselli02a,crapsi05}.
The parameters of the observed transitions were taken from the Cologne Database for Molecular Spectroscopy \citep{muller01,muller05}.
The total column density of \ce{N2D+} ($N_{\rm tot}(\ce{N2D+})$) is evaluated to be $(5.4 \pm 0.3) \times 10^{12}$ cm$^{-2}$.
The error on $N_{\rm tot}(\ce{N2D+})$ is given by propagating the errors on the total optical depth and the FWHM linewidth.  
Our $N_{\rm tot}(\ce{N2D+})$ is very close to that obtained by \citet{crapsi05} ($(4.3 \pm 0.6) \times 10^{12}$ cm$^{-2}$), 
who derived it from the \ce{N2D+}(2--1) data.

Upper limits to $N_{\rm tot}(\nzdp)$ and $N_{\rm tot}(\zndp)$ are obtained from the 3$\sigma$ upper limits to the integrated intensity
3$\sigma \sqrt{\Delta v \delta v}$ of the transitions, 
where $\sigma$ is the rms noise of the spectra, 
$\Delta v$ is the FWHM linewidth of the spectra, assumed to be the same as that of \ce{N2D+}(1--0),
and $\delta v$ is the velocity resolution.
Assuming local thermal equilibrium, the 3$\sigma$ intensity upper limits are converted into the column density upper limits by using Equation (A4) of \citet{caselli02b}, 
which is valid for optically thin emission.
$T_{\rm ex}$ is assumed to be 5 K.
We obtain $N_{\rm tot}(\nzdp) < 7.0 \times 10^9$ cm$^{-2}$ and  $N_{\rm tot}(\zndp) < 6.5 \times 10^9$ cm$^{-2}$.

\begin{table}
\caption{Derived column density and $^{14}$N/$^{15}$N ratio}
\label{table:column}
\centering
\begin{tabular}{lcc}
\hline\hline
Species & $N_{\rm tot}$ [cm$^{-2}$] & $^{14}$N/$^{15}$N \\
\hline
$\ce{N2D+}$  & $(5.4 \pm 0.3) \times 10^{12}$ & -- \\
$\nzdp$       & $<7.0 \times 10^{9}$ & $>$730 \\
$\zndp$       & $<6.5 \times 10^{9}$ &  $>$780 \\
\hline
\end{tabular}
\end{table}

\section{Discussion and Conclusion}
\label{sec:discuss}
From the column densities of the \ce{N2D+} isotopologues, we obtain the lower limits of the $^{14}$N/$^{15}$N ratio of 730 for $\nzdp$ and 780 for $\zndp$.
These lower limits are significantly larger than [$^{14}$N/$^{15}$N]$_{\rm elem}$ in the local ISM \citep[$\sim$200-300;][]{lucas98,ritchey15}.
It is reasonable to consider the $^{14}$N/$^{15}$N ratios of \ce{N2D+} as that of \ce{N2}, 
because \ce{N2D+} primary forms by \ce{N2} + \ce{X2D+}, where X is H or D.
Then our observations indicate that \ce{N2} is significantly depleted in $^{15}$N in the central part of L1544.
If we assume $T_{\rm ex}$ of 4.5 K (5.5 K) in the column density evaluation, 
the lower limits of the $^{14}$N/$^{15}$N ratios become 600 (950) for $\nzdp$ and 640 (1010) for $\zndp$.
Our qualitative conclusion is thus robust against the assumed value of $T_{\rm ex}$.
\citet{colzi18} recently derived the [$^{14}$N/$^{15}$N]$_{\rm elem}$ ratio in the local ISM of $\sim$400 
from the observations of HCN isotopologues toward a sample of 66 cores in massive star-forming regions.
Even if this higher elemental abundance ratio is adopted, our qualitative conclusion does not change.

As described in Sect. \ref{sec:intro}, nitrogen isotope exchange reactions make \ce{N2} enriched in $^{15}$N;
they are thus not relevant to the observed fractionation.
Isotope selective photodissociation of \ce{N2} by penetrating interstellar FUV photons is also ruled out as the cause of  the $^{15}$N depletion, 
because the penetration would be negligible in the central part of L1544, where \ce{N2D+} emission arises.
Our lower limits of the $^{14}$N/$^{15}$N ratios for \ce{N2D+} are consistent with those of \ce{N2H+} ($\sim$1000) obtained by \citet{bizzocchi13}, 
which also supports that isotope selective photodissociation of \ce{N2} is not responsible for the $^{15}$N depletion, as discussed in Sect. \ref{sec:intro}.
Note that cosmic ray induced photodissociation of \ce{N2} does not cause $^{15}$N fractionation, 
because the destruction of \ce{N2} by \ce{He+} is much faster \citep{heays14,furuya18a}.
Therefore, in-situ chemistry is unlikely responsible for the $^{15}$N depletion in \ce{N2} in the central part of L1544.

The most likely explanation is that the $^{15}$N depletion is inherited from more diffuse gas as recently proposed by \citet{furuya18a},
based on their astrochemical models in forming and evolving molecular clouds.
They found that during the evolution of molecular clouds, the nitrogen isotopes can be differentially partitioned between gas and ice, 
making $^{15}$N depleted gas and $^{15}$N enriched ice.
In the molecular cloud, where external UV radiation field is not fully shielded,
$^{14}$N$^{15}$N is selectively photodissociated w.r.t $^{14}$\ce{N2}, which results in the enrichment of $^{15}$N in the photodissociation product, atomic N.
Atomic N is adsorbed onto grain surfaces and converted into \ce{NH3} ice by surface reactions, 
while adsorbed \ce{N2} does not react with other species, including atomic H.
As long as the non-thermal desorption (especially photodesorption in their models) of \ce{NH3} ice is less efficient than
that of \ce{N2} ice, the net effect is the loss of $^{15}$N from the gas phase, producing $^{15}$N-depleted gas and $^{15}$N-enriched ice.
Once the external UV radiation field is sufficiently shielded, the $^{15}$N depletion does not proceed anymore 
but are largely conserved unless a significant amount of \ce{NH3} ice is sublimated.

As noted by \citet{furuya18a}, the mechanism is the most efficient around the chemical transition from atomic N to \ce{N2}, 
where the self-shielding of $^{14}$\ce{N2} becomes important.
Before the transition, both $^{14}$\ce{N2} and $^{14}$N$^{15}$N are efficiently photodissociated, 
while after the transition, the abundance of atomic N is too low to affect the bulk gas isotopic composition.
Therefore, if the mechanism proposed by \citet{furuya18a} was at work in the parent cloud of L1544 or the outer regions of L1544, 
it means that the transition from atomic to molecular nitrogen should have occurred there as well. 

Nitrogen chemistry mainly consists of three competing processes; 
(i) the conversion of atomic N to \ce{N2} in the gas phase, 
(ii) destruction of \ce{N2} via e.g. photodissociation and reaction with \ce{He+}, 
and (iii) freeze out of atomic N and \ce{N2} on to dust grains followed by surface reactions \citep[e.g.,][]{daranlot12,li13}.
The conversion of atomic N to \ce{N2} has been proposed to occur by slow neutral-neutral reactions, such as \ce{NO} + N and CN + N \citep{herbst73,daranlot12}.
According to the pseudo-time dependent gas-phase astrochemical model under dense cloud conditions (10$^4$ cm$^{-3}$, 10 K, 10 mag) by \citet{legal14},
the conversion of atomic N to \ce{N2} takes an order of Myr, depending on assumed elemental abundances.
In the gas-ice model of \citet{daranlot12}, under the similar physical conditions, 
the conversion of atomic N to \ce{N2} takes $\sim 5\times10^5$ yr, and it occurs after the significant fraction of nitrogen is frozen-out.
On the other hand,  \ce{N2} mainly forms via the reactions, \ce{NH2} + N and \ce{NH} + N 
around the transition from atomic to molecular nitrogen in the models of \citet{furuya18a} and \citet{furuya18b}, 
in which the dynamical evolution of molecular clouds are considered.
\ce{NH2} and \ce{NH} are mainly formed via photodesorption of \ce{NH3} ice, followed by photodissociation in the gas phase.
In this case, the formation rate of \ce{N2} from atomic N is, roughly speaking, similar to the freeze-out rate of atomic N.
Considering that interstellar ices, at least water ice, are already abundant in molecular clouds with relatively low line-of-sight visual
extinction \citep[e.g., $\sim$3 mag for Taurus dark clouds,][]{whittet93}, 
it may not be surprising that the transition from atomic to molecular nitrogen occurs in the parent cloud of L1544 or the outer regions of L1544.
It should be noted that the \ce{N2}-dominant region could be larger than those traced by \ce{N2H+} and \ce{NH3} emission, 
since their abundances are controlled not only by \ce{N2} but also by CO; 
the catastrophic CO freeze-out, which happens in the late stage of the interstellar ice formation 
at high densities \citep[$\gtrsim$10$^5$ cm$^{-3}$; e.g.,][]{pontoppidan06}, 
makes their abundances enhanced \citep[e.g.,][]{aikawa05}.

\begin{figure}
\resizebox{\hsize}{!}{\includegraphics{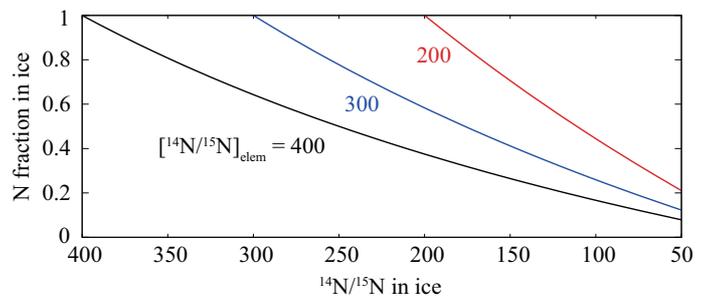}}
\caption{
Estimated fraction of elemental nitrogen in ices as functions of the $^{14}$N/$^{15}$N ratio of the ices.
The estimated fractions with different [$^{14}$N/$^{15}$N]$_{\rm elem}$ are shown by different colors. 
In all cases, the $^{14}$N/$^{15}$N ratio of the bulk gas is assumed to be 1000.
}
\label{fig:partition}
\end{figure}

One may be interested in estimating the partitioning of elemental nitrogen between gas and ice.
The abundance of gaseous \ce{N2} in dense prestellar cores was previously estimated 
from the comparison of N-chemistry models 
with the observations of \ce{N2H+} and other relevant species.
\citet{maret06} inferred that gaseous \ce{N2} contains only a few \% of overall elemental nitrogen  
\citep[N/H = $6\times10^{-5}$ in the local ISM;][]{przybilla08} in dense cloud B68.
\citet{pagani12} also inferred that gaseous \ce{N2} contains $\lesssim$1 \% of elemental nitrogen in the prestellar core L183.
While \citet{maret06} suggested that atomic N is the primary form of elemental nitrogen in B68 to account for this low gaseous \ce{N2} abundance, 
their model actually predicts that \ce{NH3} ice is the primary nitrogen reserver \citep[see also][]{daranlot12,furuya18b}.
The gas-ice astrochemical model by \citet{ruaud16}, on the other hand, predicts that the HCN ice is more abundant than \ce{NH3} ice.
Figure \ref{fig:partition} shows the fraction of elemental nitrogen in the form of ice as functions of the $^{14}$N/$^{15}$N ratio of the bulk ice.
In the figure, the $^{14}$N/$^{15}$N ratio of the bulk gas (i.e., that of \ce{N2}) is assumed to be 1000.
If \ce{NH3} and HCN ices are the primary forms of elemental nitrogen in L1544 as predicted by the astrochemical models, 
the $^{14}$N/$^{15}$N ratio of the icy species should be close to, but slightly lower than [$^{14}$N/$^{15}$N]$_{\rm elem}$.

\citet{gerin09} found that the $^{14}$N/$^{15}$N ratio of gaseous \ce{NH2D} is $>$700 in L1544.
The $^{14}$N/$^{15}$N ratios of gaseous \ce{NH3} in the cold gas of dense molecular clouds 
were derived to be 334 $\pm$ 50 in Barnard 1 and 340 $\pm$ 150 in NGC 1333 \citep{lis10}.
These measurements indicate that gaseous \ce{NH3} in the cold gas is not significantly enriched in $^{15}$N.
It should be noted, however, that the origin of  gaseous \ce{NH3} in the cold gas, i.e., whether it is formed by gas phase reactions or 
released from ices via non-thermal desorption, remains unclear.
The $^{14}$N/$^{15}$N ratio of icy species in star-forming regions 
could be constrained by measuring the molecular $^{14}$N/$^{15}$N ratios in the warm ($\gtrsim$100 K) gas surrounding protostars.
This type of observations is crucial for better understanding of the nitrogen partitioning.

Finally, the observations of comets have found that cometary \ce{NH3} and HCN are enriched in $^{15}$N by a factor of around 
three \citep{mumma11,shinnaka16} compared to the Sun \citep[$\sim$150 versus 441;][]{marty11}.
Ammonia abundance with respect to water in cometary ices (0.4-1.4 \%) is lower than that in interstellar ices (typically $\sim$5 \%) \citep[][]{mumma11,oberg11}.
These (possible) differences between the N-bearing species in cometary and interstellar ices might indicate a primordial variation in the ice formation environments
or ice processing in the solar nebula \citep[e.g.,][]{lyons09,furuya14}.

\begin{acknowledgements}
We are grateful to the IRAM staff for excellent support.
We thank Yuto Sato for his help in preparing for the IRAM proposal, 
and also thank the anonymous referee for useful comments that
helped to improve this paper.
This work is partly supported by JSPS KAKENHI Grant Numbers 16K17657 and 17K14245.
\end{acknowledgements}

\begin{appendix}
\section{}
\begin{table}
\caption{Hyperfine frequencies for $J$ = 1--0 transitions of $\nzdp$ and $\zndp$ taken from \citet{dore09}}
\label{table:freq}
\centering
\begin{tabular}{lccc}
\hline\hline
Species & $F'$--$F$ & Frequency [GHz]  & Relative intensity\\
\hline
$\nzdp$  & 1--1 &  76.0168733 & 1.000 \\
             & 2--1 &  76.0185970 & 1.667 \\
             & 0--1 &  76.0211239 & 0.333 \\
$\zndp$  & 1--1 &  74.7605619 & 1.000 \\
             & 2--1 &  74.7609788 & 1.667 \\
             & 0--1 &  74.7615650 & 0.333 \\
\hline
\end{tabular}
\end{table}
\end{appendix}


\begin{thebibliography}{}
\bibitem[Aikawa et al.(2005)]{aikawa05} Aikawa, Y., Herbst, E., Roberts, H., \& Caselli, P. 2005, \apj, 620, 330
\bibitem[Bergin \& Tafalla(2007)]{bergin07} Bergin E. A., \& Tafalla M., 2007, ARA\&A, 45, 339
\bibitem[Bizzocchi et al.(2010)]{bizzocchi10} Bizzocchi, L., Caselli, P., \& Dore, L. 2010, \aap, 510, L5
\bibitem[Bizzocchi et al.(2013)]{bizzocchi13} Bizzocchi, L., Caselli, P., Lenardo, E., \& Dore, L. 2013, \aap, 555, 109
\bibitem[Caselli et al.(2002a)]{caselli02a} Caselli, P., Walmsley, C. M., Zucconi, A., et al. 2002a, \apj, 565, 331
\bibitem[Caselli et al.(2002b)]{caselli02b} Caselli, P., Walmsley, C. M., Zucconi, A., et al. 2002b, \apj, 565, 344
\bibitem[Charnley \& Rodgers(2002)]{charnley02} Charnley, S. B., \& Rodgers, S. D. 2002, ApJL, 569, L133
\bibitem[Colzi et al.(2018)]{colzi18} Colzi, L., Fontani, F., Rivilla, V. M., et al. 2018, \mnras, 976
\bibitem[Crapsi et al.(2005)]{crapsi05} Crapsi, A., Caselli, P., Walmsley, C.M., et al. 2005, \apj, 619, 379
\bibitem[Daranlot et al.(2012)]{daranlot12} Daranlot J., Hincelin U., Bergeat A., et al. 2012, PNAS, 109, 10233
\bibitem[De Simone et al.(2018)]{desimone18} De Simone, M., Fontani, F., Codella, C., et al. 2018, \mnras, 476, 1982
\bibitem[Dore et al.(2009)]{dore09} Dore, L., Bizzocchi, L., Degli Esposti, C., \& Tinti, F. 2009, \aap, 496, 275
\bibitem[Furuya \& Aikawa(2014)]{furuya14} Furuya, K., \& Aikawa, Y. 2014, \apj, 790, 97
\bibitem[Furuya \& Aikawa(2018)]{furuya18a} Furuya, K., \& Aikawa, Y., 2018, \apj, 857, 105
\bibitem[Furuya \& Persson(2018)]{furuya18b} Furuya, K., \& Persson, M. V. 2018, \mnras, 476, 4994
\bibitem[G{\'e}rin et al.(2009)]{gerin09} G{\'e}rin, M., Marcelino, N., Biver, N., et al. 2009, \aap, 498, L9
\bibitem[Heays et al.(2014)]{heays14} Heays, A. N., Visser, R., Gredel, R., et al. 2014, \aap, 562, 61
\bibitem[Herbst \& Klemperer(1973)]{herbst73}  Herbst E., \& Klemperer W., 1973, \apj, 185, 505
\bibitem[Hily-Blant et al.(2013a)]{hilyblant13a} Hily-Blant, P., Bonal, L., Faure, A., \& Quirico, E. 2013a, Icarus, 223, 582
\bibitem[Hily-Blant et al.(2013b)]{hilyblant13b} Hily-Blant, P., Pineau des For\^ets, G., Faure, A., Le Gal, R., \& Padovani, M. 2013b, \aap, 557, 65
\bibitem[Lefloch et al.(2018)]{lefloch18} Lefloch, B., Bachiller, R., Ceccarelli, C., et al. 2018, \mnras
\bibitem[Le Gal et al.(2014)]{legal14} Le Gal, R., Hily-Blant, P., Faure, A., Pineau des For\^ets, G., Rist, C., \& Maret, S., 2014, \aap, 562, 83
\bibitem[Li et al.(2013)]{li13} Li, X., Heays, A. N., Visser, R., et al. 2013, \aap, 555, 14
\bibitem[Lis et al.(2010)]{lis10} Lis D. C., Wootten A., Gerin M., \& Roueff E., 2010, \apj, 710, L49
\bibitem[Lucas \& Liszt(1998)]{lucas98} Lucas, R., \& Liszt, H. 1998, \aap, 337, 246
\bibitem[Lyons et al.(2009)]{lyons09} Lyons, J. R., Bergin, E. A., Ciesla, F. J., et al. 2009, Geochim. Cosmochim. Acta, 73, 4998
\bibitem[Maret et al.(2006)]{maret06} Maret S., Bergin E. A., Lada C. J., 2006, Nature, 442, 425
\bibitem[Marty et al.(2011)]{marty11} Marty, B., Chaussidon, M., Wiens, R. C., Jurewicz, A. J. G., \& Burnett, D. S. 2011, Sci, 332, 1533
\bibitem[M\"uller et al.(2001)]{muller01} M\"uller, H. S. P., Thorwirth, S., Roth, D. A., \& Winnewisser, G. 2001, \aap, 370, L49
\bibitem[M\"uller et al.(2005)]{muller05} M\"uller, H. S. P., Schl\"oder, F., Stutzki, J., \& Winnewisser, G. 2005, JMoSt, 742, 215
\bibitem[Mumma \& Charnley(2011)]{mumma11} Mumma, M., \& Charnley, S. B. 2011, ARA\&A, 49, 471
\bibitem[\"Oberg et al.(2011)]{oberg11} \"Oberg, K. I., Boogert, A. C. A., Pontoppidan, K. M., et al. 2011, \apj, 740, 109
\bibitem[Pagani et al.(2012)]{pagani12} Pagani L., Bourgoin A., Lique F., 2012, \aap, 548, L4
\bibitem[Pety et al.(2005)]{pety05} Pety, J. 2005, in EDP Sciences Conf. Ser., eds. F. Casoli, T. Contini, J. Hameury, \& L. Pagani, Vol. SF2A-2005, 721
\bibitem[Pontoppidan(2006)]{pontoppidan06} Pontoppidan, K. M. 2006, \aap, 453, L47
\bibitem[Przybilla et al.(2008)]{przybilla08} Przybilla N., Nieva M.-F., Butler K., 2008, \apj, 688, L103
\bibitem[Redaelli et al.(2018)]{redaelli18} Redaelli, E., Bizzocchi, L., Caselli, P., et al. 2018, arXiv:1806.01088
\bibitem[Ritchey et al.(2015)]{ritchey15} Ritchey, A. M., Federman, S. R., \& Lambert, D. L. 2015, ApJL, 804, L3
\bibitem[Roueff et al.(2015)]{roueff15} Roueff, E., Loison, J. C., \& Hickson, K. M. 2015, \aap, 576, 99
\bibitem[Ruaud et al.(2016)]{ruaud16} Ruaud, M., Wakelam, V., \& Hersant, F. 2016, \mnras, 459, 3756
\bibitem[Shinnaka et al.(2016)]{shinnaka16} Shinnaka, Y., Kawakita, H., Jehin, E., et al. 2016, \mnras, 462, 195
\bibitem[Spezzano et al.(2016)]{spezzano16} Spezzano, S., Bizzocchi, L., Caselli, P., Harju, J., \& Brünken, S. 2016, \aap, 592, L11
\bibitem[Terzieva \& Herbst(2000)]{terzieva00} Terzieva, R., \& Herbst, E. 2000, \mnras, 317, 563
\bibitem[Whittet(1993)]{whittet93} Whittet, D. C. B. 1993, in Dust and Chemistry in Astronomy (Bristol and Philadelphia: Institute of Physics Publishing), 9
\bibitem[Wirstr\"om \& Charnley(2018)]{wirtrom18} Wirstr\"om, E. S., \& Charnley, S. B. 2018, \mnras, 474, 3720
\end{thebibliography}
\end{document}